# Towards Assessing Necessary Competence

A position statement

C. Michael Holloway
Safety-Critical Avionics Systems Branch
NASA Langley Research Center
Hampton, Virginia USA
c.michael.holloway@nasa.gov

Chris W. Johnson
Department of Computing Science
University of Glasgow
Glasgow, Scotland, UK
christopher.johnson@glasgow.ac.uk

*Abstract*—We sketch a series of studies and experiments designed to provide empirical evidence about the truth or falsity of claims that non-prescriptive approaches to standards demand greater competence from regulators than prescriptive approaches require.

I. INTRODUCTION

Common usage today divides standards into two broad categories. *Prescriptive* standards place specific requirements on development processes and practices. *Non-prescriptive* (also known as goal based, performance based, argument based, assurance case) standards place no restrictions on development processes, but require instead producing rigorous arguments that justify confidence that a system satisfies relevant properties [1][2].

Within certain domains, complying with a series of standards that tend to be more prescriptive than not is the primary means of receiving approval to release a system or vehicle to the public. In domains such as commercial aviation, this approach has been very successful. Aviation companies have highly skilled people in the necessary technical disciplines who know how to comply with, and when necessary go beyond the requirements of, the existing standards. Regulatory authorities also have highly skilled people in the necessary disciplines to assess compliance with the applicable standards (and any additional requirements that may be deemed to be appropriate for a particular system).

A common criticism of relying on prescriptive standards is that it slows down, or perhaps even prohibits, the introduction of new technologies, including technologies that may be able to enhance safety. Non-prescriptive standards, which contain no technology-specific requirements, are suggested by some to be a possible means to remove or at least lower the barrier to use of new technologies.

Moving towards non-prescriptive standards is not without critics, however. A common criticism of non-prescriptive standards is that using them will increase substantially the intellectual burden on individuals within regulatory authorities by requiring unrealistic general levels of competence. Whereas assessing compliance with a prescriptive standard only requires competence in the specific methods and processes dictated by the standard, assessing compliance with a non-prescriptive standard, the critics say, requires competence in all possible methods and processes that might be used, and competence in evaluating arguments. To do the job well, a regulator must not only be an engineer but also a philosopher or logician.

Is this criticism valid?

On first thought, it appears to be self-evidently so. Surely, a person who is assessing products that are all developed using similar processes need not know as much about the technical discipline as a person who is assessing products that may be developed using any possible collection of processes. Likewise, a person who is assessing products that all use the same basic argument – compliance with the standard – to assert acceptability need not know as much about evaluating arguments as a person who may be confronted with an unconstrained collection of arguments.

But that which is self-evidently so is not always truly so. We believe that studies and experiments should be conducted to provide empirical evidence about the truth or falsity of the claim that non-prescriptive approaches to standards demand greater competence from individual regulators than prescriptive approaches require[1]. Our first thoughts on a possible structure of these studies and experiments follow.

II. INITIAL DECISIONS

Before conducting studies or designing experiments, several decisions must be made. Among these decisions are the following:

- Whether to concentrate on one or multiple domains. Because regulatory approaches differ substantially in different domains, the results obtained in one domain may or may not be applicable in another.

- Whether to concentrate on one or multiple countries. Regulatory approaches in some domains try to be

---

[1] A reviewer rightly noted that individual competence is not as important as organizational competence within a regulatory environment. We agree; one of us (Chris) is actively involved in investigating issues related to organizational competence. We focus here on individual competence because it is the specific issue that has been raised often in regards to the practicality of non-prescriptive standards, particularly by individual regulators themselves.

consistent across national boundaries (aviation, for example); but this is not the case for other domains (rail, for example).

- Whether to concentrate on one or multiple system attributes. Studies and experiments designed to collect evidence about competency to assess if a system possesses a single attribute (safety, for example) will differ from studies and experiments with a broader scope.

- What to try first: historical literature review, surveys, prototype case study, multiple case studies, small scale experiment, or something else.

For the purposes of this position paper, we suggest that one reasonable set of decisions is the aviation domain, the US and European Union, safety, and starting with a survey followed by some small scale experiments. We suggest the aviation domain because current practices, which are based largely on prescriptive standards, have been widely successful, and assertions are commonly made that these regulatory practices inhibit innovation. Thus it is a domain with potentially high negative consequences of moving towards non-prescriptive standards prematurely, but also potentially high positive consequences of adopting such standards appropriately. We suggest the US and EU because they are responsible for the bulk of regulation within aviation. We suggest safety because ensuring adequate safety is the reason for many of the existing standards.

III. A SURVEY

The reason for starting with a survey is to collect information about the current state of practice, particularly as it is relevant to assessing current competency levels within regulatory bodies. This survey would solicit responses concerning education, post-college training, work activities including hours, knowledge about a variety of topics including reasoning and argumentation, how often certain types of knowledge are used on the job, perceived inadequacies in education and training for accomplishing necessary tasks, and undoubtedly some other relevant topics.

We recognize that the likelihood that any regulatory authority will participate in this survey is rather small. We propose it nevertheless, because designing valid and informative experiments without accurate information about the current state is difficult. Should a representative number of people respond to the survey, then a representative picture of may be drawn about what regulators currently do and what they now need to know to accomplish their jobs effectively.

IV. SMALL EXPERIMENTS

After the survey is conducted and analyzed, we propose that experiments be designed to test the following hypotheses:

- Regulators possess the requisite knowledge to assess whether a system complies with current standards. (Within the aviation domain and the regulatory authorities suggested for these experiments, this hypothesis is almost certainly true.)

- Regulators do not possess the requisite knowledge to evaluate the relative strengths of different arguments for the same conclusion.

- Regulators do not possess the requisite knowledge to assess the adequacy of an argument that a system is safe, if that argument does not rely on compliance to current prescriptive standards.

The likelihood that a statistically significant number of regulators will participate in experiments such as these is miniscule; but making the effort to find participants seems worthwhile nonetheless. If a few regulators are willing to participate, then undertaking the experiments may yield illustrative, but by no means valid conclusive, results.

Setting aside the question of whether subjects can be obtained for these experiments, considering how to design the experiments to test the suggested hypotheses is challenging. Among the challenges posed are these:

- Creating a system and associated documentation that is realistic enough for experimental validity but simple enough to allow the 'right' answers to be determined.

- Deciding how much, if any, training subjects should receive.

- Determining how to present the arguments.

- Creating the arguments so that a 'right' assessment of them exists.

- Many other things not listed above.

Of the three possible hypotheses we suggested, the second is the only one that may be feasible to simulate using non-regulators. Experiments with hypotheses related to comparative ability to evaluate argument strength have been done before, although no studies to our knowledge have specifically targeted regulators as the primary subject group.

V. CONCLUSION

We have presented some preliminary ideas designed to move towards empirically determining the truth or falsity of the claim that non-prescriptive approaches to standards demand greater competence from individual regulators than prescriptive approaches require. No matter how appealing in theory non-prescriptive standards may be, if implementing such standards in practice is likely to reduce the effectiveness of regulatory oversight, then continuing with current approaches is a wiser choice. Enhancing and then following through on the ideas we have presented will be difficult, if not altogether impossible. Conducting the survey and studies we have sketched will not answer the question definitely, but it will move the discussion away from purely speculative opinions towards evidence-based opinions.

REFERENCES

[1] J. Knight, Fundamentals of dependable computing for software engineers. Boca Raton: CRC Press, 2012.
[2] R. Hawkins, I. Habli, T. P. Kelly, J. McDermid, "Assurance cases and prescriptive software safety certification: a comparative study," Safety Science 59, pp. 55-71, November 2013